\documentclass[epj]{svjour}
\usepackage{graphicx}
\usepackage{mathtext}

\begin{document}

\title{Non-Fickian Diffusion and the Accumulation of Methane Bubbles in Deep-Water Sediments}

\author{D.~S.~Goldobin\inst{1,3} \and
 N.~V.~Brilliantov\inst{1} \and
 J.~Levesley\inst{1} \and
 M.~A.~Lovell\inst{2} \and
 C.~A.~Rochelle\inst{4} \and
 P.~D.~Jackson\inst{4} \and
 A.~M.~Haywood\inst{5} \and
 S.~J.~Hunter\inst{5} \and
 J.~G.~Rees\inst{4}}

\institute{Department of Mathematics, University of Leicester,
 Leicester LE1 7RH, UK
\and Department of Geology, University of Leicester,
 Leicester LE1 7RH, UK
\and Institute of Continuous Media Mechanics, UB RAS,
 Perm 614013, Russia
\and British Geological Survey, Keyworth,
 Nottingham NG12 5GG, UK
\and School of Earth and Environment, University of Leeds,
 Leeds LS2 9JT, UK}

\date{\today}

\abstract{
In the absence of fractures, methane bubbles in deep-water
sediments can be immovably trapped within a porous matrix by
surface tension. The dominant mechanism of transfer of gas mass
therefore becomes the diffusion of gas molecules through
porewater. The accurate description of this process requires
non-Fickian diffusion to be accounted for, including both
thermodiffusion and gravitational action. We evaluate the
diffusive flux of aqueous methane considering non-Fickian
diffusion and predict the existence of extensive bubble mass
accumulation zones within deep-water sediments. The limitation on
the hydrate deposit capacity is revealed; too weak deposits cannot
reach the base of the hydrate stability zone and form any bubbly
horizon.
}

\maketitle

\section{Introduction}
The occurrence of methane bubbles within porous water-saturated
sediments is widespread around the ocean margins. The gas within
them plays an important role in both submarine hazards, such as
submarine landslides~\cite{Kayen-Lee-1991} as well as the
formation of resources, such as  methane-hydrate deposits
\cite{Davie-Buffett-2001,Davie-Buffett-2003b,Archer-2007}.
The stability of the bubbles has a significant control on the
methane flux from the sediments into the ocean-atmosphere system.

In porous sediments the bubbles are trapped within the matrix
pores. Large moving bubbles are unstable, as they split into
smaller bubbles during migration
\cite{Lyubimov-etal-2009,Barry-etal-2010}, and smaller bubbles are trapped
by pore-throats or by surface tension forces. The minimum pore
throat diameter $l$ required to trap a small bubble, when there is
no strong pumping of the fluid through the porous matrix, can be
calculated. The surface tension forces $\sigma l$ ($\sigma$ is the
surface tension) should overwhelm the buoyancy force $\rho gl^3$
($\rho$ is the density, $g$ is the gravity), i.e.,
$l<\sqrt{\sigma/\rho g}$.  For gas-water systems, one finds,
$l<2.7\,\mathrm{mm}$. Making allowance for the inhomogeneity of
pores and the geometry of contacts, one should decrease this
estimate to $l\approx1\,\mathrm{mm}$, which still suggests
trapping even for sands. For a soft mud, mechanics of bubbles and
porous matrix can be different~\cite{Algar-Boudreau-Barry-2011}
and is not considered here.

As bubbles are trapped, and in the absence of significant
groundwater movement transporting dissolved methane, the dominant
mechanism of methane mass transfer in deep-water sediment is by
diffusion of methane through porewater, controlled by (i) the
methane saturation of the aqueous solution throughout the sediment
volume, and (ii) non-Fickian diffusion laws. The latter pertain to
two processes: firstly, the geothermal gradient causes
thermodiffusion (the Soret effect
\cite{Soret-1879}),
where the temperature gradient induces solute flux (as recognized
in several fields, e.g.\
\cite{Richter-1972});
and secondly, the impact of gravity on dissolved molecules.  These
non-Fickian contributions to the diffusive flux mean that the
solute flux cannot solely be determined by the gradient of the
solute concentration.

Currently non-Fickian diffusion is rarely considered in the
modelling of deep-water sediment systems
(\cite{Davie-Buffett-2001,Davie-Buffett-2003b,Haacke-Westbrook-Riley-2008,Garg_etal-2008},
and others involved in gas hydrate modelling only address Fickian
diffusion). However, we suggest that non-Fickian processes are
important; in particular, they may cause methane to migrate
against the direction of the steepest decrease of concentration
under certain conditions. Importance of non-Fickian diffusion was
also demonstrated for the salinity transport in hydrate-bearing
sediments \cite{Goldobin-CRM-2013}. Unfortunately, the
experimental value of the thermodiffusion coefficient for the
aqueous methane solution is unknown and can only be roughly
assessed theoretically. Therefore, we treat it as a free parameter
in our investigation. This uncertainty also justifies the fact
that prior numerical modelling work did not include non-Fickian
processes.

In researching the horizontal through-flow of water and vertical
aqueous oxygen transport in porous bubble-bearing sediments,
\cite{Donaldson-etal-1997}
considered the hydrodynamic dispersion ({\it or} ``turbulent
diffusion''
\cite{Bernard-Wilhelm-1950}),
caused by water transport through irregular pore channels, as a
transport mechanism and reasonably neglected the molecular
diffusion. Indeed, the turbulent diffusion plays a significant
role in vertical dissolved gas transport near the earth's surface.
However, it becomes insignificant away from the sediment-water
interface in deep-water sediments, where vertical and horizontal
displacements of water are comparable. Here it is significantly
smaller than molecular diffusion---even in sandy sediments.

In the present study we consider the diffusive migration of
methane in sea-floor sediments where water is saturated with
methane, and some methane is gaseous (forming bubbles; see the
sketch in Fig.\,\ref{fig1}a). In deep-water sediments the bubbly
zone is overlaid by the methane hydrate stability zone, where
non-dissolved methane forms not gaseous bubbles but hydrate, which
is included in our treatment as well.

\section{Transport processes in sediments}
On the field scale, deep-water sediments are typically much more
uniform horizontally rather than vertically. Consequently, we
consider a system that is uniform horizontally. The depth below
the water-sediment interface is measured by the $z$-coordinate
(Fig.\,\ref{fig1}).

\subsection{Diffusion in non-isothermal aqueous solutions}
Under non-isothermal conditions the diffusive flux of solute mass
is governed by the law (cf
\cite{Bird-Stewart-Lightfoot-2007,Goldobin-Brilliantov-2011})

\begin{equation}
\vec{J}_\mathrm{diff}=
 -\chi\phi\rho_\mathrm{f}D\omega\left[\frac{\nabla\omega}{\omega}
 +\alpha\frac{\nabla T}{T}-\frac{\tilde{M}\vec{g}}{RT}\right].
\label{eq-01}
\end{equation}
Here $\omega$ is the mass fraction of the solute in the solvent,
$D$ is the solute molecular diffusion coefficient, $\phi$ is the
porosity of the solid matrix, $\chi$ is the tortuosity factor
featuring the pore geometry, $\rho_\mathrm{f}$ is the fluid
density. The first term describes the ``ordinary'' Fickian
diffusion,
 $\vec{J}_\mathrm{Fick}=-\chi\phi\rho_\mathrm{f}D\nabla{\omega}$.
The second term represents the thermodiffusion effect appearing in
non-isothermal systems, where temperature inhomogeneity causes a
solute flux. The strength of the thermodiffusion effect is
characterized by the thermodiffusion constant $\alpha$ (the
conventional Soret {\it or} separation coefficient
$S_T=\alpha/T$). The third term describes the action of gravity on
solute molecules; $R=8.31\,\mathrm{J/(mol\,K)}$ is the universal
gas constant, $\tilde{M}=M^\mathrm{g}-N_1M^\mathrm{host}$,
$M^\mathrm{g}$ and $M^\mathrm{host}$ are the molar masses of the
solute and solvent, respectively, and $N_1$ is the number of
solvent molecules in the volume occupied by one solute molecule in
the solution. The value of $N_1$ can be precisely derived for
$\mathrm{CH_4}$--$\mathrm{H_2O}$ systems from the dependence of
the solution density on its concentration
\cite{Hnedkovsky-Wood-Majer-1996};
one obtains $N_1=2.23$ and $\tilde{M}=-24.3\,\mathrm{g/mol}$.

When the liquid is saturated with gas bubbles, the concentration
of solute in solvent equals the solubility, $\omega=\omega^{(0)}$,
throughout the liquid volume in the bubble-bearing zone because
the bubbles are in local thermodynamic equilibrium with the
solution. Thus, the solute flux depends merely on the temperature
and pressure fields, $T(z)$ and $P(z)$, and the solution
concentration is not a free variable,
$\omega(z)=\omega^{(0)}(T(z),P(z))$. At high pressure and low
enough temperature, the hydrate form is more thermodynamically
preferable for methane than the gaseous form. The hydrate zone can
overlie the bubbly-bearing zone (Fig.\,\ref{fig1}). In the
presence of hydrate the solution concentration equals the
solubility at equilibrium with hydrate. For the calculation of the
solubility (in bubbly and hydrate zones), half-empiric models
developed in
\cite{Duan-Mao-2006,Sun-Duan-2007} are employed.

\subsection{Pressure and temperature}
The solubility profiles and thus transport processes depend on the
pressure and temperature profiles. The system is essentially
characterized by the hydrostatic pressure $P$ and geothermal
temperature gradient $G$. Although the role of porosity
nonuniformity for the geothermal gradient was demonstrated in
\cite{Goldobin-EPL-2011}, we follow the conventional
approximation of a linear temperature profile
\cite{Davie-Buffett-2001,Garg_etal-2008,Haacke-Westbrook-Riley-2008}.
Therefore,

\begin{equation}
 P(z)=P_0+\rho_\mathrm{f}g(z+H),\qquad
 T(z)=T_\mathrm{sf}+Gz,
\label{eq-02}
\end{equation}
where $P_0$ is the atmospheric pressure, $H$ is the depth of the
water body above the bubble-bearing porous sediments, and
$T_\mathrm{sf}$ is the temperature at the sediment-water
interface.

\subsection{Sediment compaction and transport processes}
The sedimentation process and the weight of the above-laying
sediments result in a non-uniform porosity profile and a
non-uniform velocity of the downward sediment motion. The sediment
porosity is typically adopted in the form
$\phi(z)=\phi_0\exp(-z/L)$
\cite{Davie-Buffett-2001,Haacke-Westbrook-Riley-2008}, and the
resulting sediment velocity is \cite{Davie-Buffett-2001}

\begin{equation}
v_\mathrm{s}(z)=\frac{1-\phi_0}{1-\phi(z)}v_\mathrm{s}(0)\,,
\label{eq-03}
\end{equation}
where $v_\mathrm{s}(0)$ is the sedimentation rate.

The sedimentation and compaction processes create an ascending
filtration flux of water through the sediments. The net water mass
flux $\vec{J}_\mathrm{H_2O}$ is contributed by the filtration flux
$\vec{u}_\mathrm{f}$ and by the water transport in hydrate, which
moves with sediments;

\begin{equation}
\vec{J}_\mathrm{H_2O}=\rho_\mathrm{f}\vec{u}_\mathrm{f}
 +K_\mathrm{H_2O}\rho_\mathrm{h}h\phi\vec{v}_\mathrm{s}\,,
\label{eq-04}
\end{equation}
where $\rho_\mathrm{h}$ is the hydrate density,
$K_\mathrm{H_2O}\approx0.866$ is the mass fraction of
$\mathrm{H_2O}$ in methane hydrate, and $h$ is the volumetric
fraction of hydrate in pores {\it or} hydrate saturation.

Whilst the volumetric fraction of bubbles ({\it or} gas
saturation) in pores, say $b$, is vanishingly small, bubbles are
immovably trapped in pores and move with sediments
\cite{Abaci-Edwards-Whittaker-1992,Davie-Buffett-2001,Haacke-Westbrook-Riley-2008}.
However, transport processes in the system can result in a gas
saturation which exceeds the critical value $b_\mathrm{cr}$. When
the critical saturation is exceeded, gas leakage occurs.

The net methane mass flux $\vec{J}_\mathrm{CH_4}$ is contributed
by the ascending filtration flux of aqueous solution, the motion
of gas bubbles and/or hydrate with sediments, and molecular
diffusion in aqueous solution. In the absence of gas leakage, it
reads

\begin{eqnarray}
&&
\vec{J}_\mathrm{CH_4}=\rho_\mathrm{f}\omega\vec{u}_\mathrm{f}
 +(K_\mathrm{CH_4}\rho_\mathrm{h}h+\rho_\mathrm{b}b)\phi\vec{v}_\mathrm{s}
\nonumber\\
&&\qquad
 -\chi(1-h-b)\phi\rho_\mathrm{f}D\omega(\nabla\ln\omega+\beta\nabla\ln{T})
\,,
\label{eq-05}
\end{eqnarray}
where the tortuosity factor is assumed $\chi=0.75$
\cite{Haacke-Westbrook-Riley-2008},
$K_\mathrm{CH_4}=1-K_\mathrm{H_2O}$ is the mass fraction of
$\mathrm{CH_4}$ in hydrate, and

\begin{equation}
\beta=\alpha-\frac{\tilde{M}g}{RG}\,.
\label{eq-06}
\end{equation}
Eq.\,(\ref{eq-05}) with $\beta=0$ corresponds to the Fickian
diffusion law,
$\vec{J}_\mathrm{Fick}=-\chi\phi\rho_\mathrm{f}D\nabla\omega$, and
$\beta$ characterizes the {\it strength of non-Fickian part} of
the solute flux. As the value of $\alpha$ is unknown, $\beta$ is
treated as a free parameter in our study. In our model, the zones
of hydrate and gas bubbles do not overlap; either $b$ or $h$ can
be non-zero at a given location.

While the diffusion coefficient $D$ is nearly independent of
pressure~\cite{Sachs-1998}, its dependence on temperature is not
to be neglected as the temperature change from $273\,\mathrm{K}$
to $323\,\mathrm{K}$ causes the diffusivity increase by factor
$4$. The formula
 $D(T)=D_0(T/T_0)\exp[B/(T+\tau)-B/(T_0+\tau)]$
with $D_0=7.38\cdot10^{-10}\,\mathrm{m^2/s}$,
$T_0=273\,\mathrm{K}$, $B=212\,\mathrm{K}$,
$\tau=71.5\,\mathrm{K}$ well fits experimental data summarized
by~\cite{Sachs-1998}.

In the present work we do not consider details of the process of
methane generation from the organic part of sediments, assuming
that it takes place in the upper part of sediments and methane is
simply present at certain depth of about $100-200\,\mathrm{m}$
below the water-sediment interface
(cf~\cite{Haacke-Westbrook-Riley-2008}).

\section{Results}
Both the numerical modelling with finite difference method and
analytical treatment were performed. The results of numerical
simulation (also including different gas leakage models) are in
agreement with the analytical theory (as can be later seen from
Fig.\,\ref{fig2}d). It is convenient to start the presentation of
results with time-independent solutions to the problem.

\begin{figure}[t]
\center{
 \includegraphics[width=0.49\textwidth]%
 {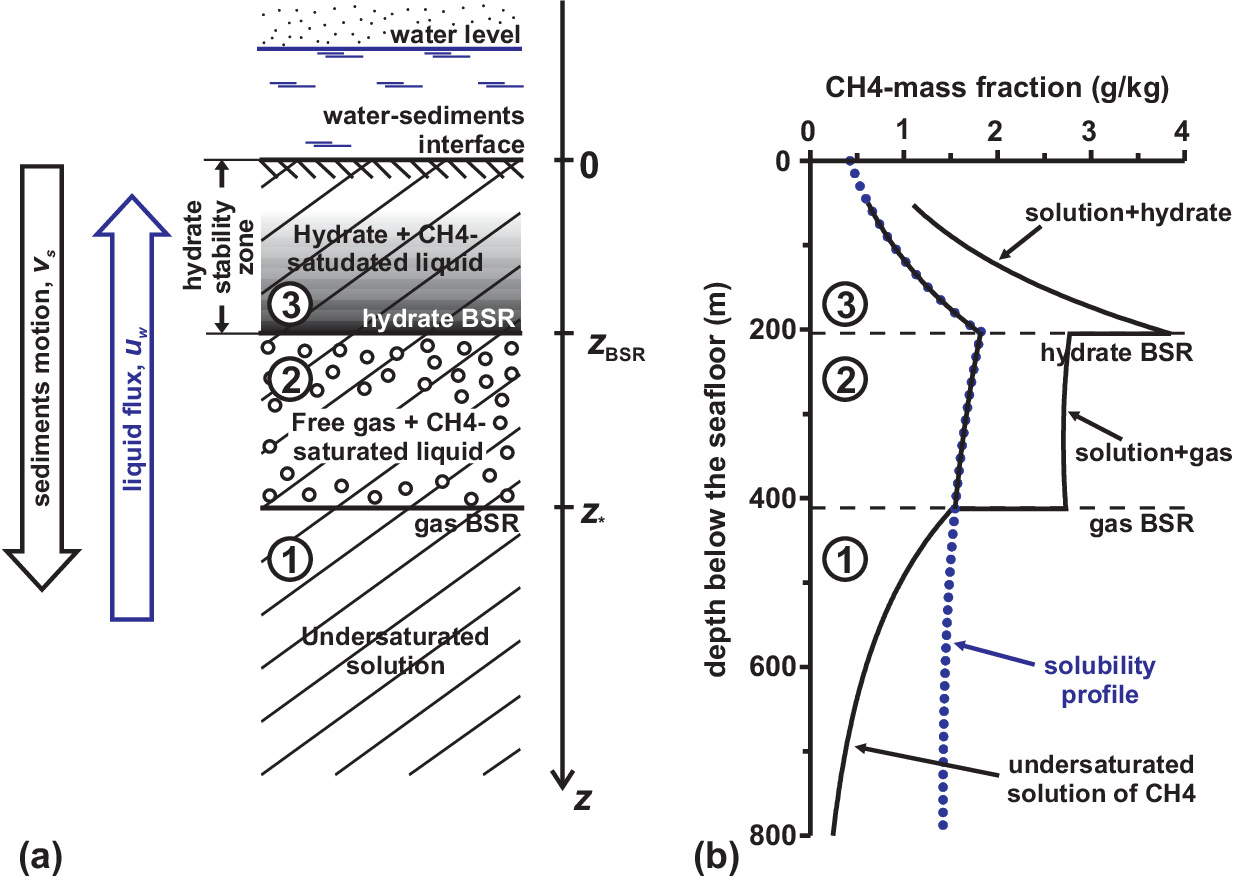}}

  \caption{
(a): Sketch of bubble-bearing marine sediments with a hydrate zone
(3) overlaying the bubbly zone (2). (b): The aqueous methane
solubility profile is plotted with the dotted blue curve for
water-body depth $H=1.5\,\mathrm{km}$ and parameters specified in
Tab.\,\ref{tab}, which correspond to the west Svalbard continental
slope \cite{Haacke-Westbrook-Riley-2008}. Black solid line plots
the net mass fraction of methane in pores---gaseous methane or
methane in hydrate are added to the mass in the aqueous
solution---for a time-independent state and $\beta=2.13$ (i.e.,
$\alpha=1.8$), see Eqs.\,(\ref{eq-07})--(\ref{eq-09}).
 }
  \label{fig1}
\end{figure}
\begin{figure*}[t]
\center{
 \includegraphics[width=0.98\textwidth]%
 {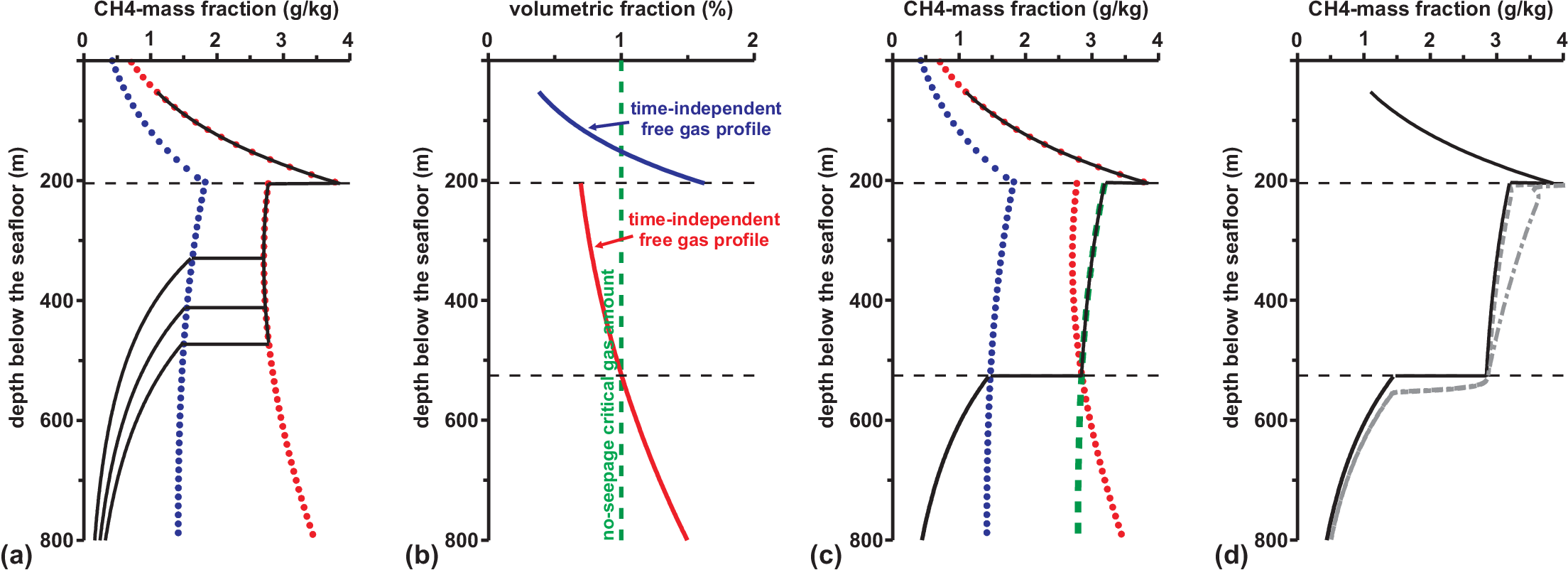}\\
}

  \caption{
Methane profiles for the parameter set specified in
Tab.\,\ref{tab}, $H=1.5\,\mathrm{km}$ and $\alpha=0$.
(a):~Time-independent methane mass profiles (cf
Fig.\,\ref{fig1}b). (b):~Volumetric fraction in pores (saturation)
of hydrate and methane for time-independent states is compared to
the critical gas saturation (green dashed line). (c):~Steady
methane mass profile with critical gas saturation at the bubbly
zone. (d):~Analytical solution profile (black solid line), and
numerically calculated steady profiles in the free gas zone with
the sediment permeability $10^{-12}\,\mathrm{m^2}$ (gray dashed
line) and $10^{-13}\,\mathrm{m^2}$ (gray dash-dotted line).
 }
  \label{fig2}
\end{figure*}
\begin{table}[b]
\caption{Geophysical properties characteristic for the west Svalbard
continental slope \cite{Haacke-Westbrook-Riley-2008}}
\begin{center}
\begin{tabular}{clc}
\hline
 & parameter & value \\
\hline
 $T_\mathrm{sf}$ & seafloor temperature &
   $-0.9^\circ\mathrm{C}$ \\
 $G$ & geothermal gradient &
   $86.5^\circ\mathrm{C/km}$\\
 $L$ & $e$-folding depth of porosity &
   $1053\,\mathrm{m}$\\
 $u_\mathrm{f0}$ & fluid filtration velocity &
   $-0.1\,\mathrm{mm/year}$\\
 $v_\mathrm{s}(0)$ & sedimentation rate &
   $0.5\,\mathrm{mm/year}$\\
 $\chi$ & tortuosity factor &
   $0.75$\\
\hline
\end{tabular}
\end{center}
\label{tab}
\end{table}

\subsection{No-leakage time-independent states}
Three zones with different transport features can be distinguished
(see Fig.\,\ref{fig1}):
\\
(1)~the lowermost zone of undersaturated aqueous solution of
methane without gas bubbles, $z>z_*$,
\\
(2)~the zone of saturated solution bearing gas bubbles,
$z_\mathrm{BSR}<z<z_*$, and
\\
(3)~the lower part of hydrate stability zone (HSZ) with saturated
solution and hydrate, $z<z_\mathrm{BSR}$.
\\
In the lower part of HSZ we consider, no process of conversion of
organic sediments into methane occurs; we assume the conversion to
be finished above this zone. In Fig.\,\ref{fig1}b, one can see a
sample profile of the net mass fraction of methane in pores
composed of dissolved methane and methane in hydrate or gas
bubbles.

At the time-independent state, neither the hydrate distribution
nor the free-gas one change with time, which requires the methane
mass flux (\ref{eq-05}) to be uniform along the sediment column.
As we do not consider methane influx from deep massifs, i.e.,
methane mass flux is zero deep in sediments, this uniform flux
should be zero everywhere. The water flux (\ref{eq-04}) is
uniform, $J_\mathrm{H_2O}=\rho_\mathrm{f}u_\mathrm{f0}$, where
$u_\mathrm{f0}$ is the filtration velocity below HSZ (cf
\cite{Davie-Buffett-2001}).

For zero methane flux and given water flux, one can find the
solution concentration in zone (1);

\begin{equation}
\omega_1(z)=\omega_1(z_*)\left[\frac{T(z_*)}{T(z)}\right]^\beta
\exp\left[\int_{z_*}^z\frac{u_\mathrm{f0}\,dz_1}{\chi\,\phi(z_1)\,D(z_1)}\right].
\label{eq-07}
\end{equation}
In zone (2), the solute concentration equals the solubility,
$\omega_2(z)=\omega_2^{(0)}(z)$, and, in the absence of leakage,
gas saturation

\begin{equation}
b=\frac{\rho_\mathrm{f}\omega_2^{(0)}}
       {\rho_\mathrm{b}v_\mathrm{s}}
 \left(-\frac{u_\mathrm{f0}}{\phi}
 +\chi D\frac{d}{dz}\ln(\omega_2^{(0)}T^\beta)\right).
\label{eq-08}
\end{equation}
In zone (3), the solute concentration equals the solubility,
$\omega_3(z)=\omega_3^{(0)}(z)$, and hydrate saturation

\begin{equation}
h=\frac{\rho_\mathrm{f}\omega_3^{(0)}}
       {K_\mathrm{CH_4}\rho_\mathrm{h}v_\mathrm{s}}
 \left(-\frac{u_\mathrm{f0}}{\phi}
 +\chi D\frac{d}{dz}\ln(\omega_3^{(0)}T^\beta)\right).
\label{eq-09}
\end{equation}
(Eqs.\,(\ref{eq-08}) and (\ref{eq-09}) are derived from
Eqs.\,(\ref{eq-04}) and (\ref{eq-05}) for small $h$ and $b$.) In
Fig.\,\ref{fig1}b the time-independent state is plotted for
$\beta=0$, the graphs for different $\beta$ are qualitatively
similar. Gas saturation $b$ immediately above the lower boundary
of the bubbly zone is finite and enough for the formation of the
second bottom simulating reflector of seismic waves, which is not
associated with hydrate.

\subsection{Evolution of the bubbly zone
            and persistence of marine hydrate deposits}
For a time-independent state determined by
Eqs.\,(\ref{eq-07})--(\ref{eq-09}) the position of the boundary
between the bubbly zone and the zone of undersaturated solution is
not imposed; all three profile plotted is Fig.\,\ref{fig2}a
satisfy the equations. Now let us consider small deviation from
these profiles.

\begin{figure}[t]
\center{
 \includegraphics[width=0.40\textwidth]%
 {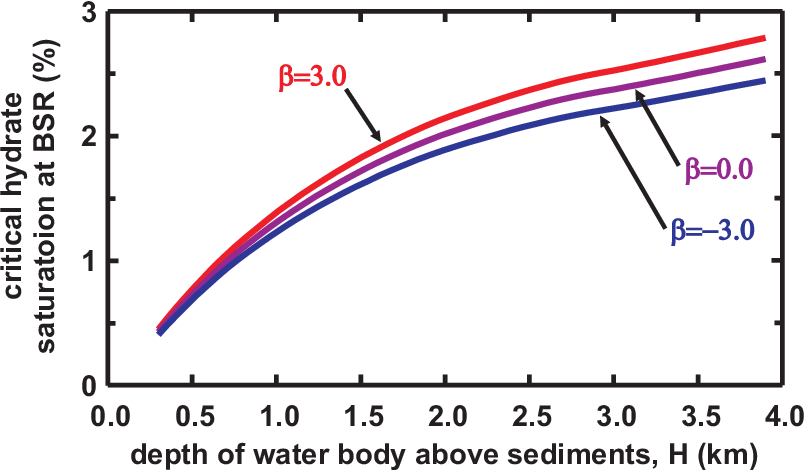}\\
}

  \caption{
Minimal hydrate saturation (volumetric fraction in pores) at the
base of the hydrate stability zone, which is required for the
hydrate deposit to be persistent, is plotted vs the water-body
depth $H$ for parameters specified in Tab.\,\ref{tab}.
 }
  \label{fig3}
\end{figure}

When hydrate saturation $h$ at the base of HSZ is smaller than
that in the time-independent state, upward methane flux appears.
Indeed, the diffusion flux is the same as in the time-independent
state, because it is controlled by the solubility profile, while
we have strictly downward transfer of hydrate with sediments, and
this transfer is diminished in comparison to the transfer in the
time-independent state. In the time-independent state these two
fluxes are balanced to zero net flux, while with the deficiency of
hydrate we have the deficiency of downward flux and thus the net
flux is upward. An upward net flux of methane results in depletion
of methane deep in sediments and gradual retreat of the bubbly
zone. After some period of time the bubbly zone disappears and,
moreover, hydrate zone retreats from the base of HSZ to the area
of methane generation from sediments. In this case one observes
HSZ with undersaturated aqueous solution and no hydrate at the
base of it and no bubbly zone; this has been reported to be
widespread in nature
\cite{Proc_ODP-V164}. Thus, one can see that the hydrate deposit
cannot be too weak, otherwise it is not persistent. In
Fig.\,\ref{fig3} the minimal hydrate saturation is plotted as a
function of the water-body depth; this value is also affected by
the non-Fickian drift strength $\beta$, for a larger $\beta$ a
stronger deposit is required.

When the hydrate saturation at the base of HSZ exceeds the
critical hydrate saturation, one observes an opposite situation: a
downward net flux of methane and a bubbly zone gradually advancing
into deep sediments. However, the advance of the bubbly zone is
not unbounded. In Fig.\,\ref{fig2}b, one can see that for
time-independent states the gas saturation increases with depth
and at certain depth, $z_*$, exceeds the critical gas saturation.
The critical gas saturation, above which gas leakage starts,
depends on many factors, including the pore geometry and the solid
matrix material \cite{Abaci-Edwards-Whittaker-1992}, and the rate
of gas release from the solution (for discussion see
\cite{Haacke-Westbrook-Riley-2008}). The critical gas saturation
$b_\mathrm{cr}$ specific for different porous massifs is about
$1\%$ (e.g., see review in
\cite{Haacke-Westbrook-Riley-2008}), and we adopt this value for
our analysis. Leakage is a hydrodynamical transport and as such is
much more efficient than the molecular diffusion transport or the
motion with sediments. Therefore, any significant excess of gas
saturation over the critical value results in a mass flux which
cannot be balances by molecular diffusion or sedimentation, and,
on the time scales of the hydrate deposit formation, gas
saturation can be only under- or nearly-critical. The bubbly zone
cannot advance deeper, as a larger gas saturation is required to
have a downward methane transfer below depth $z_*$. Leaking gas is
accumulated just above the leakage zone until the critical
saturation is reached there. Then leakage zone extends, and
finally entire bubbly zone becomes the leakage zone with gas
saturation slightly exceeding the critical saturation, as shown in
Fig.\,\ref{fig2}c. Hence, for persistent hydrate deposits
($h>h_\mathrm{cr}$) one can observe the formation of extensive
bubbly zone with critical gas saturation; the lower boundary $z_*$
of this zone is determined by condition

\begin{equation}
\left.
\frac{\rho_\mathrm{f}\omega_2^{(0)}}
       {\rho_\mathrm{b}v_\mathrm{s}}
 \left(-\frac{u_\mathrm{f0}}{\phi}
 +\chi D\frac{d}{dz}\ln(\omega_2^{(0)}T^\beta)\right)\right|_{z=z_*}
 =b_\mathrm{cr}\,.
\label{eq-10}
\end{equation}

In Fig.\,\ref{fig2}d, one can compare the analytical
time-in\-de\-pen\-dent profile to the results of numerical
simulation with a comprehensive account of processes: the rate of
methane generation by anaerobic bacteria
\[
 Q(z)=A_0\lambda[1-\phi(z)]
 \exp\left(-\lambda\int_0^z\frac{dz_1}{v_s(z_1)}\right)
\]
with $\lambda=5\cdot10^{-13}\,\mathrm{s^{-1}}$ and
$A_0=14.4\,\mathrm{kg/m^3}$ (cf~\cite{Davie-Buffett-2001}) and the
gas leakage within the bubbly zone with relative permeability for
the gas phase
$k_\mathrm{g}=b(|b-b_\mathrm{cr}|-(b-b_\mathrm{cr}))/2$~\cite{Mavko-Mukerji-Dvorkin-2009}.
The numerically calculated profile slightly deviates from the
analytical one for realistic sediment permeability and this
deviation remains non-large even for as small permeability as
$10^{-13}\,\mathrm{m^2}$. Noteworthy, the base of the free-gas
zone remains unshifted even for small permeability.

In numerical simulation the described evolution was observed with
different leakage laws, which may strongly vary from system to
system (e.g.\ \cite{Abaci-Edwards-Whittaker-1992}), and realistic
values of massif permeability. As explained and demonstrated in
Fig.\,\ref{fig2}d, the bubbly zone is tolerant to specific
features of the leakage law with given $b_\mathrm{cr}$; realistic
model parameters can only change the small excess of the gas
saturation over the critical value. In our analytical theory we do
not discuss the narrow hydrate--gas recycling zone immediately
below HSZ, where gas saturation may significantly exceed the
critical saturation. Features of this zone depend on the leakage
model which is highly uncertain without thorough knowledge of the
massif properties and is site-specific.

\begin{figure}[t]
\center{
 \includegraphics[width=0.45\textwidth]%
 {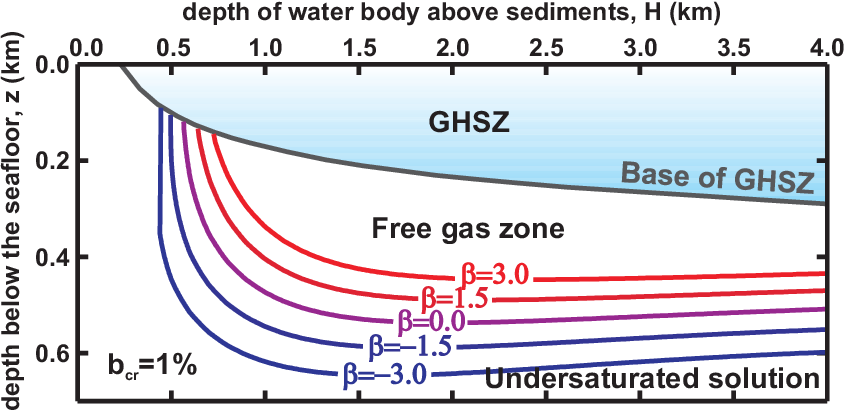}
}

  \caption{
The bubbly zone is confined between the base of the hydrate
stability zone (HSZ) and the gas bottom simulating reflector,
which is plotted in different colors for different values of the
non-Fickian drift strength $\beta$ and critical gas saturation
$b_\mathrm{cr}=1\%$. The parameter set is specified in
Tab.\,\ref{tab} and corresponds to the west Svalbard continental
slope
\cite{Haacke-Westbrook-Riley-2008}.
 }
  \label{fig4}
\end{figure}

In Fig.\,\ref{fig4}, one can see that the location of the
extensive free-gas (bubbly) zone is affected by the non-Fickian
drift strength $\beta$. Moreover, the minimal depth of the water
body above sediments needed for the extensive bubbly zone to
appear is also controlled by $\beta$. An extensive bubbly zone is
not possible below HSZ beneath shallow water bodies because the
time-independent state requires gas saturation higher than the
critical saturation and cannot be maintained because of gas
leakage. It is noteworthy that, due to hydrate--gas recycling, a
narrow bubbly layer immediately below HSZ will be still presented
wherever hydrate is present at the base of HSZ.

\subsection{Importance and uncertainties}
The behavior described is considerably influenced by the
non-Fickian drift of methane. We suggest that the Fickian
diffusion law, which has been adopted in
\cite{Davie-Buffett-2001}
and its successors and corresponds to $\beta=0$, should be
modified.

The unresolved issue here is the specific value of $\beta$ for
methane. The authors are not aware of experimental data on
thermodiffusion of methane in water, though there are a lot of
experimental studies on the thermodiffusion of methane in mixtures
of hydrocarbons (e.g.\ \cite{Wittko-Koehler-2005}). Theoretical
studies (e.g.\ \cite{Semenov-2010}) provide formulae for
calculation of the thermodiffusion constant from inter-molecular
potentials which are poorly established for water because of
hydrogen bonds. We can only calculate $\tilde{M}g/RG=-0.331$ (for
$G=86.5\,\mathrm{K/km}$) and use a rough conjecture that one can
expect $\alpha\approx1.8$ \cite{Goldobin-Brilliantov-2011}. Hence,
$\beta\approx-2.13$ can be expected for aqueous solutions with
geothermal gradient $G=86.5\,\mathrm{K/km}$.

\section{Conclusion}
We have theoretically explored the process of diffusive migration
of aqueous methane in the presence of bubbles, when they are
immovably trapped by a porous matrix---as occurs commonly in
seafloor sediments, swamps, or terrestrial aquifers. The effect of
temperature inhomogeneity across the system (geothermal gradient)
and gravitational force have been accounted for.

Non-Fickian corrections---thermodiffusion and gravitational
segregation---appear to play an important role in the migration of
methane in sediments in deep-water settings. The positive
thermodiffusion effect ($\alpha>0$, cf Eq.\,(\ref{eq-06})) and the
gravitations segregation of methane in water make negative
contribution to $\beta$, and, therefore, they assist the formation
of an extensive methane gas accumulation zone in the upper part of
the sediment column under deep water bodies. For instance,
Fig.\,\ref{fig4} illustrates that, for conditions of the west
Svalbard continental slope
\cite{Haacke-Westbrook-Riley-2008}, non-Fickian
diffusion can either extend ($\beta<0$) or shrink ($\beta>0$) the
zone of methane gas accumulation.

Remarkably, hydrate deposits with too small hydrate saturation at
the base of the hydrate stability zone should suffer diffusive
depletion and retreat from the base of HSZ to the region of
methane generation from sediments or completely disappear. This
explains why some natural hydrate deposits are reported to possess
no hydrate and no bottom simulating reflector at the base of HSZ.
The positive thermodiffusion effect and gravitational segregation,
both resulting in negative $\beta$, decrease the minimally
required hydrate saturation, as can be seen in Fig.\,\ref{fig3}.

Unfortunately, we cannot determine precise values for the
thermodiffusion constant of aqueous solutions of methane from the
literature, and can only rely on theoretical predictions (e.g.\
\cite{Semenov-2010,Goldobin-Brilliantov-2011}),
to estimate their values, as we do here. Our findings highlight
the necessity of experimental determinations of the
thermodiffusion constant for aqueous methane solutions.

\begin{acknowledgement}
The work has been supported by NERC (NE/F021941/1). DSG
acknowledges the financial support by the Government of Perm
Region (Contract C-26/212). This paper is published with the
permission of the Director of the British Geological Survey.
\end{acknowledgement}

\end{document}